\documentstyle[preprint,aps]{revtex}

\begin{document}
\draft
\title{Thermal conduction before relaxation in slowly rotating fluids}
\author{L. Herrera\thanks{
On leave from Departamento de F\'\i sica,
Facultad de Ciencias, Universidad 
Central de Venezuela, Caracas, Venezuela and Centro de Astrof\'\i sica
Te\'orica, M\'erida, Venezuela. 
}}
\address{
\'Area de F\'\i sica Te\'orica,
Facultad de Ciencias,
Universidad de Salamanca,\\
37008, Salamanca, Espa\~na.
}
\author{J. Mart\'\i nez}
\address{
Grupo de F\'{\i}sica Estad\'{\i}stica, Departamento de F\'\i sica, 
Universidad Aut\'onoma de Barcelona,\\
08193 Bellaterra, Barcelona, Espa\~na.
}

\maketitle

\begin{abstract}
For slowly rotating fluids, we establish the existence of a critical point
similar to the one found for non-rotating systems. As the fluid approaches
the critical point, the effective inertial mass of any fluid element
decreases, vanishing at that point and changing of sign beyond it. This
result implies that first order perturbative method is not always reliable
to study dissipative processes ocurring before relaxation. Physical
consequences that might follow from this effect are commented.
\end{abstract}
\pacs{04.25.-g, 05.70.Ln, 04.40.Dg, 97.60.-s}

\narrowtext
\section{Introduction}

An alternative path to the study of self-gravitating systems, which avoids
the use of numerical procedures and/or the introduction of too restrictive
simplifying assumptions, consists in perturbing the system, compeling it to
withdraw from equilibrium state. Then, evaluating it after its departure
from equilibrium, it is possible to study the tendency of the evolution of
the object. This is usually done following a first order perturbative method
which neglects cuadratic and higher terms in the perturbed quantities. This
applies whenever the relevant processes occuring in the self-gravitating
object take place on time scales which are of the order of, or smaller than,
hydrostatic time scale. In this case the quasistatic approximation fails 
\cite{KiWe94} (e.g. during the quick collapse phase preceding neutron star
formation) and the system is evaluated immediately after its departure from
equilibrium, where {\it immediately} means on a time scale of the order of
relaxation times.

Recently, it has been shown \cite{Herrera97,HeMa97,HeMa98,HeDiMa97} that,
for systems out of quasi-static approximation, a first order perturbative
theory is not always satisfactory. In fact, there exist systems for which
this method seems to be inadequate however small the perturbation is. These
ones are those for which the parameter 
\[
\alpha =\frac{\kappa T}{\tau (\rho +p)} 
\]
is close to, or beyond the so called critical point ($\alpha =1$). This
combination of the temperature $T$, the heat conduction coefficient $\kappa $%
, the relaxation time $\tau $, the energy density $\rho ,$ and the pressure $%
p,$ has been found to be the same in spherically symmetric systems \cite
{Herrera97,HeMa97}, and axially symmetric systems with reflection symmetry 
\cite{HeDiMa97}. Also the viscous spherically symmetric case has been
studied with similar results \cite{HeMa98}.

The astrophysical interest of the study of relativistic rotating fluids is
past all doubt. Therefore, it seems important to establish, for such
systems, the existence, or not, of a critical point as described above. With
this aim, we assume that, initially, a non-viscous slowly rotating object is
close to hydrostatic equilibrium (along the $r$ coordinate) and nearly to
thermal adjustment (the so called {\it complete equilibrium} \cite[p. 66]
{KiWe94}), as measured by a local Minkowskian observer. Therefore, the time
derivatives of the radial velocity and heat flow can be neglected. At that
time, we perturb the radial velocity and the heat flow, and we evaluate
conservation equations and heat transport equation just after the
perturbation takes place, neglecting cuadratic and higher terms in the
perturbed quantities. Here{\it \ just after the perturbation} means on a
time scale which is of the order of the relaxation time. This is necessary
if the relevant processes take place on time scales which are of the order,
or smaller than, hydrostatic time scale. This meaning of {\it just after the
perturbation }implies that physical quantities remain unchanged, but not the
time derivatives of the perturbed quantities. These ones, are still small,
but they cannot be neglected since the system is departing from the complete
equilibrium.

As has been mentioned above, it is necessary to use a heat transport
equation, together with the conservation equations, to find out the
existence, or not, of the critical point. In order to keep clear of
inconsistences, the heat transport equation cannot be the well-known Eckart
one \cite{Eckart40,Landau59} because it assumes a vanishing relaxation time.
Furthermore, this theory suffers from two importants drawbacks:
Non-causality (the thermal signals propagate at infinite speed), and
unstability (all the predicted equilibrium states are unstable).
Fortunately, there exist well physically founded thermodynamical theories
that avoid these problems and that can deal with pre-relaxation processes 
\cite{Israel76,IsSt79,PaJoCa82,JoCaLe93}. In this work, we shall use the
Israel-Stewart heat transport equation. Nevertheless, it is important to
emphasize that, as in \cite{Herrera97,HeMa97,HeMa98,HeDiMa97}, the results
found are also valid in the context of the {\it Extended Irreversible
Thermodynamics }\cite{PaJoCa82,JoCaLe93}.

The paper is organized as follows. The next section is devoted to introduce
the interior and exterior metrics used, and to construct the stress-energy
tensor. Also the validity of the slow rotating limit is discussed. In
section three, the conservation equations and heat transport equation are
evaluated just after perturbation and we find the expression for the
critical point. Finally, we discuss the results in the last section.

We adopt metric of signature $-2$ and geometrised units $c=G=1$. The
quantities subscripted with $r_1$ denote that they are evaluated at the
surface of the object, whereas a partial derivative with respect time is
denoted by subscript $,0.$

\section{Energy-momentum tensor}

We consider a nonstatic and axisymmetric distribution of matter and
radiation. Let us assume that the interior metric is given by \cite{HeJi82} 
\[
ds^2=Y^2du^2+2\frac YXdudr+2a\sin ^2\theta \left( \frac YX-Y^2\right)
dud\phi -2a\sin ^2\theta \left( \frac YX\right) drd\phi 
\]
\begin{equation}
-R^2d\theta ^2-\sin ^2\theta \left[ r^2+a^2\cos ^2\theta +2a^2\sin ^2\theta
\left( \frac YX-\frac{Y^2}2\right) \right] d\phi^2 ,  \label{metricin}
\end{equation}
where $u=x^0$ is a timelike coordinate, $r=x^1$ is the null coordinate and $%
\theta =x^2$ and $\phi =x^3$ are the usual angle coordinates. Here $%
R^2=r^2+a^2\cos ^2\theta $, $a$ is the angular momentum per unit mass in the
weak field limit -the Kerr parameter-, and $X$ and $Y$ are arbitrary
functions of $u$, $r$ and $\theta $. The $u$-coordinate is related to
retarded time in a flat space-time and therefore, $u$-constant surfaces are
null cones open to the future. In these coordinates $r$-constant surfaces
are oblate spheroids.

The energy-momentum tensor may be expressed in the above coordinates (\ref
{metricin}). Nevertheless, the physical quantities appearing in it will be
those measured by a local Minkowskian observer comoving with the fluid.
Thus, it is necessary to introduce the local Minkowski coordinates ($t$, $x$%
, $y$, $z$) related to these ones by 
\begin{eqnarray}
dt &=&Ydu+\frac{dr}X+a\sin ^2\theta \left( \frac 1X-Y\right) d\phi
\label{transform} \\
dx &=&\frac{dr}X+\frac{a\sin ^2\theta }Xd\phi  \nonumber \\
dy &=&\left[ r^2+a^2\cos ^2\theta \right] ^{1/2}d\theta  \nonumber \\
dz &=&\left[ r^2+a^2\cos ^2\theta \right] ^{1/2}\sin \theta d\phi . 
\nonumber
\end{eqnarray}

The radial velocity of matter is given by 
\begin{equation}
\frac{dr}{du}=\frac{XYR\omega _x-Ya\omega _z\sin \theta }{R\left( 1-\omega
_x\right) +Ya\omega _z\sin \theta },  \label{velocity}
\end{equation}
and the orbital velocity is 
\begin{equation}
\Omega =\frac{d\phi }{du}=\frac{Y\omega _z}{R\sin \theta \left( 1-\omega
_x\right) +Ya\omega _z\sin ^2\theta },  \label{Omega}
\end{equation}
where $\omega _x$ and $\omega _z$ are the corresponding components of the
velocity of a fluid element as measured the locally Minkowski frame.

In the slow rotating limit $a<<1,$ and consequently $\Omega <<1.$ Thus, from
(\ref{Omega}) 
\[
\omega _z=\frac{\Omega R\sin \theta \left( 1-\omega _x\right) }{Y\left(
1-\Omega a\sin ^2\theta \right) }=\frac{\Omega r\sin \theta \left( 1-\omega
_x\right) }Y+{\cal O}(\Omega ^2), 
\]
and $\omega _z$ is also much less than unity. Note that in the static case (%
{\it i.e. }$\omega _x=0$) and for a local Minkowskian observer ({\it i.e. }$%
Y=1$), this means that every fluid element must move at non-relativistic
velocity \cite{Hartle67}. A simple calculus shows that this condition is
accomplished by most of the known pulsars \cite[p. 146]{Demianski85}.

The interior metric (\ref{metricin}), can be matched to the Kerr-Vaidya
exterior metric \cite{CaKa77} 
\[
ds^2=\left( 1-\frac{2mr}{\widetilde{R}^2}\right) du^2+2dudr+\frac{4mr%
\widetilde{a}\sin ^2\theta }{\widetilde{R}^2}dud\phi -2\widetilde{a}\sin
^2\theta drd\phi 
\]
\begin{equation}
-\widetilde{R}^2d\theta ^2-\sin ^2\theta \left[ r^2+\widetilde{a}^2+\frac{2mr%
\widetilde{a}^2\sin ^2\theta }{\widetilde{R}^2}\right] d\phi ,
\label{metricout}
\end{equation}
where $\widetilde{R}=r^2+\widetilde{a}^2\cos ^2\theta ,$ $\widetilde{a}$ is
the exterior Kerr parameter and $m$ is the total mass. It is worth
mentioning at this point that the metric above is not a pure radiation
solution and may be interpreted as such only asymptotically \cite{GoHeJi79}.
A pure rotating radiation solution may be found in reference \cite{KrHa95}.
However, although the interpretation of the Carmeli-Kaye metric is not
completely clear, the appearance of the critical point is independent of the
shape and the intensity of the emission pulse, and will be put in evidence
for small values of luminosity (see below).

A particular solution can be found in \cite{HeJi82}. Nevertheless, in this
work we shall not restrict ourselves to a particular solution, and we shall
go on using the unknown functions $X(u,r,\theta )$ and $Y(u,r,\theta ).$
These ones are constrained by the following junction conditions at the
surface ($r=r_1$) \cite{HeJi82} 
\begin{eqnarray}
X_{r=r_1} &=&Y_{r=r_1}=\left( 1-\frac{2mr}{R^2}\right) _{r=r_1}^{1/2}, 
\nonumber \\
\left( \frac{\partial X}{\partial \theta }\right) _{r=r_1} &=&\left( \frac{%
\partial Y}{\partial \theta }\right) _{r=r_1}=-\left( \frac{mra^2\sin
(2\theta )}{X\left( r^2+a^2\cos ^2\theta \right) ^2}\right) _{r=r_1},
\label{constrain} \\
\left( \frac{\partial X}{\partial r}\right) _{r=r_1} &=&\left( \frac{%
\partial Y}{\partial r}\right) _{r=r_1}=-\left( \frac{m\left( r^4-a^4\cos
^2\theta +2r^2a^2\cos ^2\theta \right) }{X\left( r^2-a^2\cos ^2\theta
\right) \left( r^2+a^2\cos ^2\theta \right) ^2}\right) _{r=r_1}.  \nonumber
\end{eqnarray}

Next, we assume that, for a local Minkowskian observer, comoving with the
fluid, the space-time contains:

\begin{enumerate}
\item  An anisotropic fluid of density $\rho ^{mat},$ radial pressure $%
p^{mat}$ and tangential pressure $p_{\perp }^{mat}.$

\item  A radiation field of specific intensity $I(x,t;\vec{n},\nu )$,
radiation energy flow $q$, radiation energy density $\rho ^{rad}$, and
radiation pressure $p^{rad}.$
\end{enumerate}

The specific intensity of the radiation field $I(x,t;\vec{n},\nu ),$ is
measured at the position $x$ and time $t,$ traveling in the direction $\vec{n%
}$ with a frequency $\nu $. The moments of $I(x,t;\vec{n},\nu )$ for a
planar geometry can be written as \cite{Mihalas84} 
\begin{equation}
\rho ^{rad}=\frac 12\int_0^\infty d\nu \hspace{.25cm}\int_{-1}^1d\mu %
\hspace{.25cm}I(x,t;\vec{n},\nu )\hspace{.25cm},  \label{eq0mom}
\end{equation}
\begin{equation}
q=\frac 12\int_0^\infty d\nu \hspace{.25cm}\int_{-1}^1d\mu \hspace{.25cm}\mu
I(x,t;\vec{n},\nu )\hspace{.25cm}  \label{eq1mom}
\end{equation}
and 
\begin{equation}
p^{rad}=\frac 12\int_0^\infty d\nu \hspace{.25cm}\int_{-1}^1d\mu %
\hspace{.25cm}\mu ^2I(x,t;\vec{n},\nu )\hspace{.25cm}.  \label{eq2mom}
\end{equation}
where $\mu =\cos \theta $. In classical radiative transfer theory, the
specific intensity of the radiation field, $I(x,t;\vec{n},\nu )$ at the
position $x$ and time $t$, traveling in the direction $\vec{n}$ with a
frequency $\nu $, is defined so that, 
\begin{equation}
d{\cal E}=I(x,t;\vec{n},\nu )\;dS\;\cos \alpha d\vartheta \;d\nu \;dt,
\end{equation}
is the energy crossing a surface element $dS$, into solid angle $d\vartheta $
around $\vec{n}$ ($\alpha $ is the angle between $\vec{n}$ and the normal to 
$dS$), transported by radiation of frequencies ($\nu ,\nu +d\nu $), in time $%
dt$ (see \cite{Mihalas84} for details).

For a nonrotating observer, the radiation portion of the stress-energy
tensor reads \cite{Mihalas84,Lindquist66} 
\begin{equation}
\hat{T}_{\mu \nu }^R=\left( 
\begin{array}{cccc}
\rho ^{rad} & -q & 0 & 0 \\ 
-q & p^{rad} & 0 & 0 \\ 
0 & 0 & ~\frac 12(\rho ^{rad}-p^{rad}) & 0 \\ 
0 & 0 & 0 & ~~\frac 12(\rho ^{rad}-p^{rad})
\end{array}
\right) ,  \label{tradmin}
\end{equation}

The radiation part of the energy momentum tensor as seen by an observer
comoving with the fluid can be found by means of a local rotation to (\ref
{tradmin}) 
\begin{equation}
\hat{T}_{\mu \nu }^R=\left( 
\begin{array}{cccc}
\rho ^{rad}+D^2p_{\perp }^{rad} & -q & 0 & DG \\ 
-q & p^{rad} & 0 & -Dq \\ 
0 & 0 & p_{\perp }^{rad} & 0 \\ 
DG & -Dq & 0 & D^2\rho ^{rad}+p_{\perp }^{rad}~
\end{array}
\right) ,  \label{tradminrot}
\end{equation}
where $D$ is an unknown function of $u$, $r$ and $\theta $ associated with
the local dragging of inertial frames effect, $p_{\perp }^{rad}=~\frac
12(\rho ^{rad}-p^{rad})$ and $G=\frac 12{\cal (}3\rho ^{rad}-p^{rad}).$

The material part of the energy-momentum tensor for this observer is given
by 
\begin{equation}
\hat{T}_{\mu \nu }^M=(\rho _M+p_{\perp })\widehat{U}_\mu \widehat{U}_\nu
-p_{\perp }\eta _{\mu \nu }+(p-p_{\perp })\widehat{s}_\mu \widehat{s}_\nu ,
\label{tmatminrot}
\end{equation}
where the Minkowski metric is denoted by $\eta _{\mu \nu }$, $\widehat{s}%
_\mu =\delta _\mu ^x$ and $\widehat{U}_\mu =\delta _\mu ^t.$ Thus, the
energy-momentum tensor, as seen by a Minkowskian observer comoving with the
fluid, can be written as 
\begin{equation}
\widehat{T}_{\mu \nu }=\hat{T}_{\mu \nu }^R+\hat{T}_{\mu \nu }^M.
\label{tttt}
\end{equation}
In the slow rotation limit $D$ is taken up to first order. Thus, in virtue
of (\ref{tradminrot}) and (\ref{tmatminrot}), (\ref{tttt}) $\widehat{T}_{\mu
\nu }$ can be expressed as 
\begin{equation}
\widehat{T}_{\mu \nu }=(\rho +p_{\perp })\widehat{U}_\mu \widehat{U}_\nu
-P_{\perp }\eta _{\mu \nu }+(p-p_{\perp })\widehat{s}_\mu \widehat{s}_\nu +2%
\widehat{q}_{(\mu }\widehat{U}_{\nu )}+2\widehat{q}_{(\mu }\widehat{D}_{\nu
)}+2G\widehat{U}_{(\mu }\widehat{D}_{\nu )},  \label{tmunumin}
\end{equation}
where $\widehat{q}_\mu =-q\widehat{s}_\mu $, $\widehat{D}_\mu =D\delta _\mu
^z$, $\rho =\rho ^{rad}+\rho ^{mat}$ is the total energy density, and $%
p=p^{mat}+p^{rad}$ and $p_{\perp }=p_{\perp }^{mat}+p_{\perp }^{rad}$ are
the total radial pressure and the total tangential pressure respectively.

It remains to express the energy-momentum tensor in curvilinear coordinates (%
\ref{metricin}), as seen by an observer at rest with respect to the
Minkowskian coordinates given by (\ref{transform}). Thus, we apply a Lorentz
boost and the coordinate transformation defined in (\ref{transform}). The
boost velocity is, in the rotating case,{\bf \ }$\vec{\omega}=(\omega
_x,0,\omega _z)$ -see \cite{HeMeNuPa94} for details. Assuming slow rotation
limit, $D$, $a$ and $\omega _z$ are taken up to first order. Thus, 
\begin{equation}
T_{\mu \nu }=(\rho +p_{\perp })U_\mu U_\nu -p_{\perp }g_{\mu \nu
}+(p-p_{\perp })s_\mu s_\nu +2q_{(\mu }U_{\nu )}+2q_{(\mu }D_{\nu
)}+2GU_{(\mu }D_{\nu )},  \label{tmunubon}
\end{equation}
where, $g_{\mu \nu }$ is given by (\ref{metricin}), 
\begin{equation}
U_\mu =\gamma Y\delta _\mu ^u+\ \frac{\gamma \left( 1-\omega _x\right) }%
X\delta _\mu ^r+\gamma \left[ \ a\sin ^2\theta \left( \frac{1-\omega _x}%
X-Y\right) -\omega _zr\sin \theta +{\cal O}(\omega _z^2)\right] \delta _\mu
^\phi ,  \label{umu}
\end{equation}
\[
s_\mu =-\gamma \omega _xY\delta _\mu ^u+\left[ \ \frac{\gamma (1-\omega _x)}%
X+{\cal O}(\omega _z^2)\right] \delta _\mu ^r 
\]
\begin{equation}
+\left[ r\sin \theta \frac{\omega _z}{\omega _x}(\gamma -1)+\frac{\gamma
a\sin ^2\theta }X\left[ 1-\omega _x\left( 1-YX\right) \right] +{\cal O}%
(\omega _z^2)\right] \delta _\mu ^\phi  \label{smu}
\end{equation}
\begin{equation}
q_\mu =-qs_\mu ,  \label{qmu}
\end{equation}
\begin{equation}
D_\mu ={\cal O}(\omega _z^2)\delta _\mu ^u+{\cal O}(\omega _z^2)\delta _\mu
^r+\ \left[ Dr\sin \theta +{\cal O}(\omega _z^2)\right] \delta _\mu ^\phi ,
\label{dmu}
\end{equation}
\[
\omega =\sqrt{\omega _x^2+\omega _z^2}=\omega _x+{\cal O}(\omega _z^2), 
\]
and 
\begin{equation}
\gamma =\frac 1{\sqrt{1-\omega ^2}}=\frac 1{\sqrt{1-\omega _x^2}}+{\cal O}%
(\omega _z^2).  \label{gamma}
\end{equation}
Here, ${\cal O}(\omega _z^2)$ corresponds to cuadratic and higher terms in $%
\omega _z$, $D$ and $a$.

\section{Departure from complete equilibrium}

As has been mentioned in the Introduction, we assume that, before
perturbation, the slowly rotating system is evolving along a sequence of
states in which it is close complete equilibrium. Therefore, $u$-derivatives
of $\omega _x$ and $q$ can be neglected because it is close to hydrostatic
equilibrium (along the $r$ coordiante) and nearly thermally adjusted. A
system which is thermally adjusted changes its properties considerabily only
within a time scale $\tau _{cha}$ that is large as compared with the
Kelvin-Helmholtz time scale $\tau _{KH}$ \cite[p. 66]{KiWe94}. Thus, before
perturbation we can assume that the $u$-derivatives of $\rho $, $p$ and $%
p_{\perp }$ are small, and consequently $\omega _x$ too ({\it i.e.} we can
neglect cuadratic and higher terms in $\omega _x$). On the other hand, if
the system is close to hydrostatic equilibrium, then the hydrostatic time
scale $\tau _{hyd}\sim \sqrt{r^3/m}$ is much shorter than the
Kelvin-Helmholtz time scale $\tau _{KH}\sim m^2/2rl,$ and inertial terms in
the equation of motion $T_{r;\mu }^\mu =0$ can be ignored. This condition
will be accomplished for small values of luminosity $l$, and consequently
for small values of $q.$ Thus, before perturbation $q\sim {\cal O}(\omega
_x).$

We shall evaluate the system just after perturbation. Where, as stated
before, {\it just after perturbation} means on a time scale of the order of
the relaxation time. Physically, this implies that the perturbed quantities (%
$\omega _x$ and $q$) are still much less than unity. Nevertheless, the
system is departing form complete equilibrium and the $u$-derivatives of $%
\omega _x$ and $q$ must be small but different from zero ({\it i.e.} $%
q_{,0}\sim \omega _{x,0}\sim {\cal O}(\omega _x)$).

Thus, the system is characterized by:

\begin{itemize}
\item  Before perturbation 
\begin{equation}
\rho _{,0}\sim p_{,0}\sim p_{\perp ,0}\sim \omega _x\sim q\sim {\cal O}%
(\omega _x)
\end{equation}
\begin{equation}
\omega _{x,0}\sim q_{,0}\sim {\cal O}(\omega _x^2).
\end{equation}

\item  After perturbation 
\begin{equation}
\rho _{,0}\sim p_{,0}\sim p_{\perp ,0}\sim \omega _x\sim q\sim \omega
_{x,0}\sim q_{,0}\sim {\cal O}(\omega _x)
\end{equation}
\end{itemize}

In order to clarify the existence of a critical point in slowly rotating
fluids, we shall use conservation equations ($T_{\nu ;\mu }^\mu =0$).

\subsection{Conservation equations\label{conservation}}

Before perturbation, conservation equations read 
\begin{equation}
R_\nu :=T_{\nu ;\mu }^\mu =0.  \label{before}
\end{equation}

After perturbation, physical quantities contained in (\ref{tmunubon}) remain
unchanged since we are evaluating the system on a time scale of the order of
the relaxation time. Therefore, the only new terms appearing in conservation
equations are those containing $u$-derivatives of $\omega _x$ and $q,$ and
conservation equations can be written as 
\begin{equation}
\widetilde{T}_{\nu ;\mu }^\mu =\widetilde{R}_\nu +\widetilde{\omega }_{x,0}%
{\cal F}_\nu +\widetilde{q}_{,0}{\cal G}_\nu =0,  \label{after}
\end{equation}
where tilde denotes that the quantity is evaluated after perturbation, and $%
{\cal F}_\nu $ and ${\cal G}_\nu $ do not depend on $\omega _x,$ $q$ or $u$%
-derivatives of physical variables since we are applying first order
perturbation theory. The only terms that can contain $\widetilde{\omega }%
_{x,0}$ and $\widetilde{q}_{,0}$ in $\widetilde{T}_{\nu ;\mu }^\mu =0$ are
of the form $\widetilde{T}_{\nu ,0}^0.$ By means of (\ref{metricin}), (\ref
{umu}-\ref{dmu}) and (\ref{tmunubon}), we find four equations of the form (%
\ref{after}) -see appendix A for details- 
\begin{eqnarray}
\widetilde{R}_u &=&\left( \rho +p\right) \widetilde{\omega }_{x,0}+%
\widetilde{q}_{,0}  \label{forces} \\
\widetilde{R}_r &=&\frac 2{XY}\left[ \left( \rho +p\right) \widetilde{\omega 
}_{x,0}+\widetilde{q}_{,0}\right]  \nonumber \\
\widetilde{R}_\theta &=&0  \nonumber \\
\widetilde{R}_\phi &=&0.  \nonumber
\end{eqnarray}
Note that $\widetilde{R}_\phi $ is not the total meridional force acting on
a given fluid element since it contains terms in $\widetilde{\omega }_{z,0}. 
$ Nevertheless, $\widetilde{R}_r$ does not contain $u$-derivatives of
physical variables. Thus, $\widetilde{R}_r>0$ is the total outward force
(pressure gradient + gravitational) along the $r$-coordinate acting on a
given fluid element after perturbation.

The $u$-derivative of the heat flow $\widetilde{q}_{,0}$ can be connected
with $\widetilde{\omega }_{x,0}$ by means of an adequate heat transport
equation.

\subsection{Heat transport equation}

As it is well-known, Eckart-Landau transport equation \cite
{Eckart40,Landau59} assumes a vanishing relaxation time. This fact leads to
undesirable predictions: An infinite speed for the propagation of the
thermal signals and unstable equilibrium states \cite{HiLi83}. Thus, it is
necessary to adopt a relativistic thermodynamic theory leading to a
hyperbolic equation for the propagation of thermal signals. On the other
hand, we are evaluating the system {\it just after} its departure form
hydrostatic equilibrium and thermal adjustment (in the sense described
above). Thus, to be consistent with this choice we must use a heat transport
equation with non vanishing relaxation time.

We shall use the Israel-Stewart relativistic transport equation \cite
{Israel76,IsSt79}. For viscous free fluid distributions, this one can be
written as \cite{MaTr97a} 
\begin{equation}
\tau h^{\mu \nu }U^\alpha q_{\nu ;\alpha }+q^\mu =\kappa h^{\mu \nu }\left(
T_{,\nu }-TU^\alpha U_{\nu ;\alpha }\right) -\frac 12\kappa T^2\left( \frac
\tau {\kappa T^2}U^\beta \right) _{;\beta }q^\mu +\tau \omega ^{\mu \nu
}q_\nu ,  \label{trequa}
\end{equation}
where $\kappa ,$ $\tau $ and $T$ denote thermal conductivity, thermal
relaxation time and temperature respectively, $h^{\mu \nu }=U^\mu U^\nu
-g^{\mu \nu }$ is the projector onto the hypersurface ortogonal to $U^\mu $
and $\omega _{\mu \nu }=h_\mu ^\alpha h_\nu ^\beta U_{[\alpha ;\beta ]}$ is
the vorticity.

Before perturbation, transport equations (\ref{trequa}) can be symbolized as 
\begin{equation}
{\cal H}^\mu =0.  \label{hache}
\end{equation}

Just after perturbation (\ref{trequa}) can be written as 
\begin{equation}
\widetilde{{\cal H}}^\mu +\widetilde{\omega }_{x,0}{\cal I}^\mu +\widetilde{q%
}_{,0}{\cal J}^\mu =0.  \label{hachetilde}
\end{equation}
Nevertheless, physical quantities contained on ${\cal H}^\mu $ do not change
just after perturbation, and $\widetilde{{\cal H}}^\mu ={\cal H}^\mu $.
Thus, from (\ref{hache}), expresion (\ref{hachetilde}) takes the form 
\begin{equation}
\widetilde{\omega }_{x,0}{\cal I}^\mu +\widetilde{q}_{,0}{\cal J}^\mu =0.
\label{heatfin}
\end{equation}
Vectors ${\cal I}^\mu $ and ${\cal J}^\mu $, as ${\cal F}_\nu $ and ${\cal G}%
_\nu $ in the preceding section, do not depend on $\omega _x,$ $q$ or $u$%
-derivatives of physical variables.

The components of (\ref{trequa}) containig $u$-derivatives of $q$ and $%
\omega _x$ up to first order are 
\begin{equation}
\tau h^{\mu \nu }U^\alpha q_{\nu ;\alpha }  \label{unocov}
\end{equation}
and 
\begin{equation}
-\kappa Th^{\mu \nu }U^\alpha U_{\nu ;\alpha }.  \label{doscov}
\end{equation}
Therefore, heat transport equation (\ref{trequa}) just after perturbation is
given by - see appendix B for details - 
\begin{equation}
\widetilde{q}_{,0}=-\frac{\kappa T}\tau \widetilde{\omega }_{x,0},
\label{q0}
\end{equation}
for any value of $\mu .$

\subsection{Equation of motion}

We are now in position to find the equation of motion just after
perturbation. From (\ref{forces}) and (\ref{q0}) we can write 
\begin{equation}
\widetilde{R}_r=\frac{2\left( \rho +p\right) }{XY}\left( 1-\alpha \right) 
\widetilde{\omega }_{x,0},  \label{rr}
\end{equation}
where 
\begin{equation}
\alpha =\frac{\kappa T}{\tau (\rho +p)}.  \label{alfa}
\end{equation}
As it has been noted in section \ref{conservation}, $\widetilde{R}_r>0$ is
the total outward force along the $r$-coordinate acting on a given fluid
element. Note that it vanishes for $\alpha =1$ (the critical point). This
fact has an important consequence: For $\alpha =1,$ $\widetilde{R}_r$
vanishes even though the $u$-derivative of the radial velocity is different
from zero. This method also predicts an anomalous behaviour beyond the
critical point -equation (\ref{rr}). If $\alpha >1,$ then an outward force ($%
\widetilde{R}_r>0$) implies an inward acceleration ($\widetilde{\omega }%
_{x,0}<0$). Therefore, it seems that, for systems out of quasi-static
approximation, first order perturbation theory can not be applied close to
the critical point or beyond it.

On the other hand, (\ref{rr}) may be compared with the Newtonian form 
\begin{equation}
Force=mass\times acceleration,  \label{newton}
\end{equation}
where here, the term 
\begin{equation}
\frac{2\left( \rho +p\right) }{XY}\left( 1-\alpha \right)  \label{masainer}
\end{equation}
stands for the {\it effective inertial mass}. Below the critical point, this 
{\it inertial mass }decreases as $\alpha $ grows up. This seems to be
connected with the dynamical stability of the system, leading to a minimum
stability for $\alpha =1$ \cite{HeDi97}.

\section{Conclusions}

In this paper we have studied the departure, of a slowly rotating fluid
distribution, from a state close to hydrostatic equilibrium (along the $r$
coordinate) and nearly thermally adjusted. Our aim has been to elucidate the
existence of a critical point similar to the found for non-rotating systems 
\cite{Herrera97,HeMa97,HeMa98,HeDiMa97}. The existence of this critical
point implies that first order perturbative method is not always
satisfactory to study pre-relaxation processes ({\it i.e. }processes that
take place on time scales smaller than the hydrostatic time scale).

We have found that, also in this case, there exists such critical point.
This one is given by condition 
\begin{equation}
\alpha =\frac{\kappa T}{\tau (\rho +p)}=1,  \label{final}
\end{equation}
and it coincides with this one found in spherically simmetric case \cite
{Herrera97,HeMa97,HeMa98} and in axially simmetric case \cite{HeDiMa97}.
Therefore, condition $\alpha =1$, establishes an upper limit for which
pre-relaxation processes can be studied by means of a first order
perturbative method. This result is also valid if the initial system
configuration is strictly in complete equilibrium and radially static ({\it %
i.e. }$\omega _x=\omega _{x,0}=q=q_{,0}=0$).

Note that this method predicts, for values of $\alpha $ less than unity,
that the effective inertial mass decreases as $\alpha $ grows. Intuitively,
this means that the departure from equilibrium or quasi-equilibrium will be
steeper for larger $\alpha$'s, or, in other words, that the smaller $\alpha
, $ the larger dynamical stability. This point, has been recently
illustrated, by means of an expression for the active gravitational mass in
terms of $\alpha$ \cite{HeDi97}. It is interesting to emphasize that, at
least in non-rotating configurations, causality and stability conditions 
\cite{HiLi83} not always forbid the existence of the critical point \cite
{HeMa97,HeMa98}.

Finally it is also worth noticing that the critical point and the
inflationary equation of state for non-dissipative systems ($p = - \rho$)
are similar in that they imply the vanishing of the inertial mass term.
Therefore one might wonder about the plausibility of an inflationary
scenario in a Universe at, or close to, the critical point.

\acknowledgments

This work has been partially supported by the Spanish Ministry of Education
under Grant No. PB94-0718

\appendix

\section{Conservation equations just after perturbation}

In the slow rotating limit, $U^\mu ,$ $s^\mu $ and $D^\mu $ read 
\begin{equation}
U^\mu =\left[ \frac{\gamma \left( 1-\omega _x\right) }Y+{\cal O}(\omega
_z^2)\right] \delta _u^\mu +\left[ \omega _xX+{\cal O}(\omega _z^2)\right]
\delta _r^\mu +\left[ \frac{\omega _z}{r\sin \theta }+{\cal O}(\omega
_z^2)\right] \delta _\phi ^\mu  \label{usup}
\end{equation}
\begin{eqnarray}
s^\mu &=&\left[ \frac{\gamma \left( 1-\omega _x\right) }Y+{\cal O}(\omega
_z^2)\right] \delta _u^\mu +\left[ -\gamma X+{\cal O}(\omega _z^2)\right]
\delta _r^\mu  \label{ssup} \\
&&+\left[ -\frac{\omega _z}{r\sin \theta }\left( \frac{\gamma -1}{\omega _x}%
\right) +{\cal O}(\omega _z^2)\right] \delta _\phi ^\mu  \nonumber
\end{eqnarray}
\begin{equation}
D^\mu ={\cal O}(\omega _z^2)\delta _u^\mu +{\cal O}(\omega _z^2)\delta
_r^\mu +\left[ -\frac D{r\sin \theta }+{\cal O}(\omega _z^2)\right] \delta
_\phi ^\mu ,  \label{dsup}
\end{equation}
and $q^\mu =-qs^\mu .$ From (\ref{tmunubon}) the only terms that contain $u$%
-derivatives of $\omega _x$ and $q$ (up to first order in $\omega _x$ and $%
\omega _z$) in conservation equations $T_{\nu ;\mu }^\mu =0$ are of the form 
$T_{\nu ,0}^0.$ In particular 
\begin{eqnarray}
\left( \rho +p_{\perp }\right) \left( U^0U_\nu \right) _{,0} &=&\left( \rho
+p_{\perp }\right) \left[ \frac{\gamma \left( 1-\omega _x\right) }YU_\nu
\right] _{,0}  \label{parciales} \\
\left( p-p_{\perp }\right) (s^0s_\nu )_{,0} &=&\left( p-p_{\perp }\right)
\left[ \frac{\gamma \left( 1-\omega _x\right) }Ys_\nu \right] _{,0} 
\nonumber \\
G\left( U^0D_\nu +U_\nu D^0\right) _{,0} &=&G\left[ \frac{\gamma \left(
1-\omega _x\right) }YD_\nu \right] _{,0}  \nonumber \\
\left( q^0U_\nu +q^0D_\nu +q_\nu U^0+q_\nu D^0\right) _{,0} &=&-q_{,0}\left[ 
\frac{\gamma \left( 1-\omega _x\right) }Y\left( U_\nu +D_\nu +s_\nu \right)
\right]  \nonumber \\
&&-q\left[ \frac{\gamma \left( 1-\omega _x\right) }Y\left( U_\nu +D_\nu
+s_\nu \right) \right] _{,0}.  \nonumber
\end{eqnarray}
Thus, from (\ref{umu}-\ref{dmu}), (\ref{gamma}), (\ref{parciales}) and
following the definition of ${\cal F}_\nu $ and ${\cal G}_\nu $ given in (%
\ref{after}), we find up to first order in $\omega _x$ 
\begin{eqnarray}
\omega _{x,0}{\cal F}_u &=&\left( \rho +p_{\perp }\right) \left[ \frac
1{1+\omega _x}\right] _{,0}-\left( p-p_{\perp }\right) \left[ \frac{\omega _x%
}{1+\omega _x}\right] _{,0}-q\left[ \frac{1-\omega _x}{1+\omega _x}\right]
_{,0}  \label{F} \\
&=&-\omega _{x,0}\left( \rho +p\right) ,  \nonumber \\
\omega _{x,0}{\cal F}_r &=&\left( \frac{\rho +p-2q}{XY}\right) \left[ \frac{%
1-\omega _x}{1+\omega _x}\right] _{,0}=-\omega _{x,0}\frac{2\left( \rho
+p\right) }{XY},  \nonumber \\
\omega _{x,0}{\cal F}_\theta &=&0,  \nonumber \\
\omega _{x,0}{\cal F}_\phi &=&0,  \nonumber
\end{eqnarray}
\begin{eqnarray}
q_{,0}{\cal G}_u &=&-q_{,0},  \label{G} \\
q_{,0}{\cal G}_r &=&-q_{,0}\frac 2{XY},  \nonumber \\
q_{,0}{\cal G}_\theta &=&0,  \nonumber \\
q_{,0}{\cal G}_\phi &=&0,  \nonumber
\end{eqnarray}
where we have neglected in $\omega _{x,0}{\cal F}_\phi $ terms of the form $%
\omega _{x,0}D$, $\omega _{x,0}a$ and $\omega _{x,0}\omega _z,$ and in $%
q_{,0}{\cal G}_\phi $ terms of the form $q_{,0}D$, $q_{,0}a$ and $%
q_{,0}\omega _z.$ Thus, 
\begin{equation}
{\cal F}_\mu =-\left( \rho +P\right) \left( 
\begin{array}{llll}
1, & 2/XY, & 0, & 0
\end{array}
\right) ,  \label{Ffin}
\end{equation}
and 
\begin{equation}
{\cal G}_\mu =-\left( 
\begin{array}{llll}
1, & 2/XY, & 0, & 0
\end{array}
\right) .  \label{Gfin}
\end{equation}
The expression of conservation equations can be found by means of (\ref
{after}): 
\begin{equation}
\widetilde{R}_\nu =\left[ \left( \rho +p\right) \widetilde{\omega }_{x,0}+%
\widetilde{q}_{,0}\right] \left( 
\begin{array}{llll}
1, & 2/XY, & 0, & 0
\end{array}
\right) .  \label{constot}
\end{equation}

\section{Heat transport equation just after perturbation}

The right-hand terms in (\ref{trequa})

\[
-\frac 12\kappa T^2\left( \frac \tau {\kappa T^2}U^\beta \right) _{;\beta
}q^\alpha , 
\]
and 
\[
\tau \omega ^{\mu \nu }q_\nu , 
\]
contain factors of the form $\omega _{x,0}q,$ which are of second order.
Therefore, the only terms in (\ref{trequa}) that contain $u$-derivatives of $%
\omega _x$ and $q$ up to first order are of the form 
\begin{equation}
\tau h^{\mu \nu }U^0q_{\nu ,0}  \label{uno}
\end{equation}
and 
\begin{equation}
-\kappa Th^{\mu \nu }U^0U_{\nu ,0}.  \label{dos}
\end{equation}
Using (\ref{usup}), (\ref{smu}) and (\ref{qmu}) in (\ref{uno}) and (\ref{dos}%
) 
\begin{eqnarray}
\tau h^{\mu \nu }U^0q_{\nu ,0} &=&\frac{\gamma \left( 1-\omega _x\right) }%
Y\tau h^{\mu \nu }\left( -q_{,0}s_\nu -qs_{\nu ,0}\right)  \label{tres} \\
-\kappa Th^{\mu \nu }U^0U_{\nu ,0} &=&-\frac{\gamma \left( 1-\omega
_x\right) }Y\kappa Th^{\mu \nu }U_{\nu ,0}.  \nonumber
\end{eqnarray}
Vectors ${\cal I}^\mu {\cal \ }$and ${\cal J}^\mu $ (\ref{heatfin}) do not
depend on $\omega _x$, $q$ and their $u$-derivatives because of we are using
first order perturbation theory. Thus, from (\ref{tres}) 
\begin{equation}
\widetilde{\omega }_{x,0}{\cal I}^\mu =\widetilde{\omega }_{x,0}\frac{\kappa
T}{XY}h^{\mu r},  \label{cuatro}
\end{equation}
and 
\begin{equation}
\widetilde{q}_{,0}{\cal J}^\mu =\widetilde{q}_{,0}\frac \tau {XY}h^{\mu r},
\label{cinco}
\end{equation}
where we have neglected terms of the form $\widetilde{q}_{,0}a.$ Therefore,
from (\ref{cuatro}) and (\ref{cinco}), expresion (\ref{heatfin}) takes the
form 
\begin{equation}
\widetilde{q}_{,0}=-\frac{\kappa T}\tau \widetilde{\omega }_{x,0},
\end{equation}
which is valid for any $\mu .$

\end{document}